\documentclass[oneside]{elsart}
\usepackage{relsize}
\usepackage{amsmath}
\usepackage{graphicx}
\usepackage{amssymb}

\makeatletter

\DeclareRobustCommand{\lyxmathsym}[1]{\ifmmode\begingroup\def\b@ld{bold}
  \def\rmorbf##1{\ifx\math@version\b@ld\textbf{##1}\else\textrm{##1}\fi}
  \mathchoice{\hbox{\rmorbf{#1}}}{\hbox{\rmorbf{#1}}}
  {\hbox{\smaller[2]\rmorbf{#1}}}{\hbox{\smaller[3]\rmorbf{#1}}}
  \endgroup\else#1\fi}

\makeatother

\usepackage[english]{babel}

\begin{document}
\selectlanguage{english}
\begin{frontmatter}

\title{Geometry of the Central Limit Theorem in the Nonextensive Case}

\author{C. Vignat$^{1}$ and A. Plastino$^{2}$}

\address{$^{1}$I.G.M., Universit\'{e} de Marne la Vall\'{e}e, Marne la Vall\'{e}ee, France}

\address{$^{2}$Exact Sci. Fac., National University La Plata and IFLP-CCT-CONICET
\\
 C.C. 727, 1900 La Plata, Argentina }

\ead{vignat@univ-mlv.fr, plastino@uolsinectis.com.ar}
\begin{abstract}
We uncover geometric aspects that underlie the sum of two independent
stochastic variables when both are governed by $q-$Gaussian probability
distributions. The pertinent discussion is given in terms of random
vectors uniformly distributed on a $p-$sphere.
\end{abstract}
\end{frontmatter}

\section{Introduction}

Nonextensive statistical physics provides a rich framework for the
interpretation of complex systems' behavior whenever classical statistical
physics fails \cite{uno}. The basic tool for this approach is the
extension of the classical Boltzmann entropy to the wider class of
Tsallis entropies. In this context, the usual Gaussian distributions
is extended to the q-Gaussian distributions, to be defined below.
The study of the properties of these distributions is an interesting
problem, being the subject of a number of recent publications \cite{uno}.
Of special interest is the extension of the usual stability result
that holds in the Gaussian case, namely, that if $X_{1}\in\mathbb{R}$
and $X_{2}\in\mathbb{R}$ are independent Gaussian random variables
with unit variance, then the linear combination\[
Z=a_{1}X_{1}+a_{2}X_{2}\]
 is again Gaussian and\begin{equation}
Z\sim\sqrt{a_{1}^{2}+a_{2}^{2}}X,\label{eq:Gaussiancase}\end{equation}
 where $X$ is Gaussian with unit variance, and  $\sim$ 
denotes equality in distribution.

This stability property is at the core of the central limit theorem
(CLT), which describes the behavior of systems that result of the
additive superposition of many independent phenomena. The CLT can
be ranked among the most important results in probability theory and
statistics, and plays an essential role in several disciplines, notably
in statistical mechanics. Pioneers like A. de Moivre, P.S. de Laplace,
S.D. Poisson, and C.F. Gauss have shown that the Gaussian distribution
is the attractor of the superposition process of independent systems
with a finite second moment. Distinguished authors like Chebyshev,
Markov, Liapounov, Feller, Lindeberg, and L\'{e}vy have also made essential
contributions to the CLT-development. As far as physics is concerned
one can state that, starting from any system with any finite variance
distribution function (for some measurable quantity $x$), and combining
additively a sufficiently large number of such independent systems together, the resultant distribution function of $x$ is always Gaussian.

A natural question is thus the extension of the stability result (\ref{eq:Gaussiancase})
to the nonextensive case, that is, for q-Gaussian distributions. This
interesting problem is currently the subject of several publications
(see for example \cite{umarov}) in which possible extensions of the
CLT to the nonextensive context are studied. The aim of this communication
is to give some geometric insight into the behavior of q-Gaussian
distributions for the case $q<1$.

\section{Definitions and notations}

In nonextensive statistics, the usual Shannon entropy of a density
probability $f_{X}$, namely \[
H_{1}\left(X\right)=-\int f_{X}\log f_{X}\]
 is replaced by its Tsallis version\[
H_{q}\left(X\right)=\frac{1}{1-q}\left(1-\int f_{X}^{q}\right)\]
 where the nonextensivity index $q$ is a real parameter, usually
taken to be positive. It can be checked by applying  L'Hospital's rule that Shannon's entropy
coincides with the limit case \[
\lim_{q\to1}H_{q}\left(X\right)=H_{1}\left(X\right)\]

It is a well-known result that the distribution that maximizes the
Shannon entropy under a covariance matrix constraint $EXX^{T}=K$
(where $K$ is a symmetric definite positive matrix) is the Gaussian
distribution\[
f_{X}\left(X\right)=\frac{1}{\vert\pi K\vert^{\frac{1}{2}}}\exp\left(-X^{T}K^{-1}X\right).\]
 Its nonextensive counterpart, called a q-Gaussian, is defined as
follows.
\begin{defn}
The $n-$variate distribution with zero mean and given covariance
matrix $EXX^{T}=K$ having maximum Tsallis entropy is denoted as $G_{q}\left(K\right)$
and defined as follows for $0<q<1:$

\begin{equation}
f_{X}\left(X\right)=A_{q}\left(1-X^{t}\Sigma^{-1}X\right)_{+}^{\frac{1}{1-q}},\label{eq:q<1gaussian}\end{equation}
 with matrix $\Sigma=pK$, parameter $p$ defined as $p=2\frac{2-q}{1-q}+n$
and notation $\left(x\right)_{+}=\max\left(x,0\right).$ Moreover,
the partition function is \[
A_{q}=\frac{\Gamma\left(\frac{2-q}{q-1}+\frac{n}{2}\right)}{\Gamma\left(\frac{2-q}{1-q}\right)\vert\pi\Sigma\vert^{1/2}}.\]
 We note that this distribution has bounded support; namely, $f_{X}\left(X\right)\ne0$ only
when $X$ belongs the ellipso\"id
\[
\mathcal{E}_{\Sigma}=\left\{ Z\in\mathbb{R}^{n}\,;\, Z^{t}\Sigma^{-1}Z\le1\right\} .
\]
\end{defn}

We also need the notion of spherical vector, defined as follows:
\begin{defn}
A random vector $X\in\mathbb{R}^{n}$ is spherical if its density
$f_{X}$ is a function of the norm $\vert X\vert$ of $X$ only, namely\[
f_{X}\left(X\right)=g\left(\vert X\vert\right)\]
 for some function $g:\mathbb{R}^{+}\to\mathbb{R}^{+}.$
\end{defn}
An alternative characterization of a spherical vector is as follows
\cite{Fang}:
\begin{prop}
A random vector $X\in\mathbb{R}^{n}$ is spherical if\[
X\sim AX\]
 for any orthogonal matrix $A,$ where sign $\sim$ denotes equality
in distribution.
\end{prop}
This property highlights the importance of spherical vectors in physics
since they describe systems that are invariant by orthogonal transformation.

A fundamental property of a spherical vector is the following:
\begin{prop}
\cite{Fang} If $X\in\mathbb{R}^{n}$ is a spherical random vector,
then it has the stochastic representation \[
X\sim rU\]
 where $U$ is a uniform vector on the sphere $\mathcal{S}_{n}=\left\{ X\in\mathbb{R}^{n}; \, X^{T}X=1\right\} $
and $r$ is a positive scalar random variable independent of $U.$
Moreover, $r$ has stochastic representation \begin{equation}
r\sim\vert X\vert.\label{eq:rdistrib}\end{equation}

\end{prop}

\section{A heuristic approach}

We start with a heuristic approach to the stability problem, namely
the behavior of the random variable $Z=a_{1}X_{1}+a_{2}X_{2}$ when
$X_{1}$ and $X_{2}$ are two unit variance, q-Gaussian independent
random vectors in $\mathbb{R}^{n}$ with nonextensivity parameter
$q<1;$ let us assume that the following hypothesis - called (H) hypothesis :

\begin{equation}
n+\frac{2}{1-q}\in\mathbb{N},\label{eq:hypothesisH}\end{equation}
holds so that $\frac{1}{1-q}=\frac{p-n}{2}-1$ where $p>n$ is an
integer; a classical result is that $X_{1}$ (resp. $X_{2}$) can
then be considered as the $n-$dimensional marginal vector of a random
vector $U_{1}$ (resp. $U_{2}$) that is uniformly distributed on
the unit sphere $\mathcal{S}_{p-1}$ in $\mathbb{R}^{p}.$ Thus, there
exist random vectors $\tilde{X}_{1}$ and $\tilde{X}_{2}$ in $\mathbb{R}^{p-n}$
such that \[
U_{1}=\left[\begin{array}{c}
X_{1}\\
\tilde{X}_{1}\end{array}\right]\,\,\text{and}\,\, U_{2}=\left[\begin{array}{c}
X_{2}\\
\tilde{X}_{2}\end{array}\right]\]
 are two $p-$dimensional independent vectors uniformly distributed
on $\mathcal{S}_{p}.$ Then, the sum $U_{1}+U_{2}$ is a spherical
vector and has stochastic representation\[
a_{1}U_{1}+a_{2}U_{2}\sim rU\]
 where $U$ is uniform on $\mathcal{S}_{p}.$ Now, by equation (\ref{eq:rdistrib}),
the random variable $r$ is distributed as \[
r\sim\vert a_{1}U_{1}+a_{2}U_{2}\vert=\sqrt{a_{1}^{2}+a_{2}^{2}+2\lambda a_{1}a_{2}}\]
 where $\lambda=U_{1}^{T}U_{2}:$ this can be easily deduced from\[
\vert a_{1}U_{1}+a_{2}U_{2}\vert=\sqrt{a_{1}^{2}U_{1}^{T}U_{1}+a_{2}^{2}U_{2}^{T}U_{2}+2a_{1}a_{2}U_{1}^{T}U_{2}}\]
remarking that $U_{1}^{T}U_{1}=U_{2}^{T}U_{2}=1.$ But $\lambda$
is a random variable with q-Gaussian distribution! We prove this result
by noticing that, conditioned to $U_{2}=u_{2},$ random variable $\lambda$
is the angle between $U_{1}$ and the fixed direction $u_{2}.$ Since
$U_{1}$ is spherical, we may restrict our attention to the angle
between $U_{1}$ and the first vector of the canonical basis in $\mathbb{R}^{n},$
so that we look for the distribution of the first component of $U_{1}$,
which follows a q-Gaussian distribution with parameter $q_{\lambda}$
such that \[
\frac{1}{1-q_{\lambda}}=\frac{p-1}{2}-1.\]
Since this distribution does not depend on our initial choice $U_{2}=u_{2},$ random variable $\lambda$ follows unconditionally the above cited distribution.
We conclude that the $n-$dimensional marginal $Z=a_{1}X_{1}+a_{2}X_{2}$
of vector $a_{1}U_{1}+a_{2}U_{2}$ is distributed as\[
a_{1}X_{1}+a_{2}X_{2}\sim rX\]
 where $X$ is the $n-$dimensional marginal vector of $U$ so that
$X$ is again $q-$Gaussian with parameter $q.$ Moreover, this result
extends to the case where $X_{1}$ and $X_{2}$ both have%
\footnote{the case where $X_{1}$ and $X_{2}$ have distinct covariance matrices
is more difficult and left to further study%
} a covariance matrix $K\ne I$ by multiplying vectors $X_{1}$ and
$X_{2}$ by matrix $K^{\frac{1}{2}}.$ Consequently, we have deduced
the following
\begin{thm}
\label{thm:sum}If $X_{1}$ and $X_{2}$ are two q-Gaussian independent
random vectors in $\mathbb{R}^{n}$ with covariance matrix $K$ and
nonextensivity parameter $q<1$ and if hypothesis (H) holds then \[
a_{1}X_{1}+a_{2}X_{2}\sim\left(a_{1}\circ a_{2}\right)X\]
 where $X$ is again q-Gaussian with same covariance matrix $K$ and
same nonextensive parameter $q$ as $X_{1}$, and where\begin{equation}
a_{1}\circ a_{2}=\sqrt{a_{1}^{2}+a_{2}^{2}+2\lambda a_{1}a_{2}},\label{eq:algebra}
\end{equation}
the random variable $\lambda$ being independent of $X$ and again
q-Gaussian distributed with nonextensive parameter $q_{\lambda}$
defined by
\begin{equation}
\label{eq:qq'}
q_{\lambda}=\frac{\left(n-1\right)-\left(n-3\right)q}{\left(n+1\right)-\left(n-1\right)q}.
\end{equation}
\end{thm}

Two remarks are of interest at this point:
\begin{itemize}
\item the univariate framework $n=1$ is the only case for which random
variable $\lambda$ has the same nonextensivity parameter $q_{\lambda}$
as $X_{1}$ and $X_{2}$;
\item however, we note that 
\[
\lim_{n\to+\infty}q_{\lambda}=1.
\]
This means that for large dimensional systems, the random variable $\lambda$ converges to the constant $0$ and we recover the deterministic convolution; this is coherent with the fact that large dimensional $q-$Gaussian vectors are "close" to Gaussian vectors by De-Finetti inequality.
\end{itemize}

\begin{figure}
\label{figure2}
\begin{centering}
\includegraphics[scale=0.4]{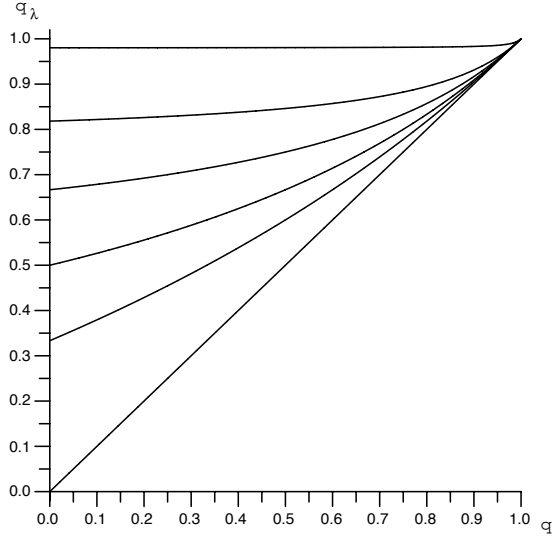}
\par\end{centering}
\caption{nonextensivity parameter $q_{\lambda}$ as a function of $q$ for
dimensions $n=1,2,3,5,10$ and $100$ (bottom to top)}
\end{figure}

The curves in Figure \ref{figure2} show the nonextensive parameter $q_{\lambda}$
as a function of $q$ for several values of dimension $n.$

More can be said about the algebra $a_{1}\circ a_{2}$:
\begin{thm}
\label{thm:associative}The algebra $a_{1}\circ a_{2}$ defined as
in (\ref{eq:algebra}) is associative and for any $n\ge2,$\[
a_{1}\circ a_{2}\circ...\circ a_{n}=\sqrt{\sum_{i=1}^{n}a_{i}^{2}+2\sum_{i<j}\lambda_{ij}a_{i}a_{j}}\]
 where random variables $\lambda_{ij}=U_{i}^{T}U_{j}$ are q-Gaussian.

As an example,\[
a_{1}\circ a_{2}\circ a_{3}=\sqrt{a_{1}^{2}+a_{2}^{2}+a_{3}^{2}+2\lambda_{12}a_{1}a_{2}+2\lambda_{13}a_{1}a_{3}+2\lambda_{23}a_{2}a_{3}}.\]
\end{thm}
\begin{pf}
By definition, \begin{eqnarray*}
a_{1}\circ a_{2}\circ...\circ a_{n} & = & \vert\sum_{i=1}^{n}a_{i}U_{i}\vert\\
 & = & \sqrt{\sum_{i=1}^{n}a_{i}^{2}U_{i}^{t}U_{i}+2\sum_{i<j}a_{i}a_{j}U_{i}^{t}U_{j}}\end{eqnarray*}
 Since $\vert U_{i}\vert=1$, we deduce, by denoting $U_{i}^{t}U_{j}=\lambda_{ij}$,
that \[
a_{1}\circ a_{2}\circ...\circ a_{n}=\sqrt{\sum_{i=1}^{n}a_{i}^{2}+2\sum_{i<j}\lambda_{ij}a_{i}a_{j}}.\]
By the same proof as above, we deduce that each $\lambda_{i}$ is
$q-$Gaussian distributed with parameter $q_{\lambda}.$ We remark that random
variables $\lambda_{i,j}$ are independent pairwise but are obviously
not mutually independent.
\end{pf}

\section{Generalization}

The preceding result was derived under the hypothesis (H) as expressed
by (\ref{eq:hypothesisH}), that is, for specific values of $q<1$
only; we show in this section that this result holds in fact without this hypothesis
- for all values of $q<1$ - but the proof requires more elaborate
analytic tools. Our main result is
\begin{thm}
\label{thm:generalsum}Theorem \ref{thm:sum} holds for all values
of $q$ such that $0<q<1.$\end{thm}
\begin{pf}
The characteristic function associated to the $q-$Gaussian distribution
(\ref{eq:q<1gaussian}) is\[
\varphi_{X}(u)\overset{d}{=}E\exp\left(iu^{T}X\right)=2^{\frac{p}{2}-1}\Gamma\left(\frac{p}{2}\right)\frac{J_{\frac{p}{2}-1}\left(\sqrt{u^{T}Ku}\right)}{\left(\sqrt{u^{T}Ku}\right)^{\frac{p}{2}-1}}\]
where $J_{\frac{p}{2}-1}$ is the Bessel function of the first kind
and with parameter $\frac{p}{2}-1$ where \[
p=2\frac{2-q}{1-q}+n.\]
According to Gegenbauer \cite[367, eq.16]{Watson},\[
2^{\nu}\Gamma\left(\nu+\frac{1}{2}\right)\Gamma\left(\frac{1}{2}\right)\frac{J_{\nu}\left(Z\right)}{Z^{\nu}}\frac{J_{\nu}\left(z\right)}{z^{\nu}}=\int_{0}^{\pi}\frac{J_{\nu}\left(\sqrt{Z^{2}+z^{2}-2Zz\cos\phi}\right)}{\left(Z^{2}+z^{2}-2Zz\cos\phi\right)^{\frac{\nu}{2}}}\sin^{2\nu}\phi d\phi.\]
 Choosing $Z=a_{1}\sqrt{u^{T}Ku}$, $z=a_{2}\sqrt{u^{T}Ku},$ $\lambda=-\cos\phi$
and $\nu=\frac{p}{2}-1,$ this equality can be rewritten as \[
\varphi_{a_{1}X_{1}}\left(u\right)\varphi_{a_{2}X_{2}}\left(u\right)=\varphi_{\sqrt{a_{1}^{2}+a_{2}^{2}+2\lambda a_{1}a_{2}}X}\left(u\right)\]
 where $\lambda$ is distributed according to \[
f\left(\lambda\right)=\frac{\Gamma\left(\nu+1\right)}{\Gamma\left(\nu+\frac{1}{2}\right)\Gamma\left(\frac{1}{2}\right)}\left(1-\lambda^{2}\right)^{\nu-\frac{1}{2}}.\]
Since $q_{\lambda}$ is defined by \[
\frac{1}{1-q_{\lambda}}=\nu-\frac{1}{2}=\frac{p-1}{2}-1,\]
we deduce (\ref{eq:qq'}).
\end{pf}
Let us recall the scaling behavior of Gaussian vectors\[
a_{1}X_{1}+a_{2}X_{2}\sim\sqrt{a_{1}^{2}+a_{2}^{2}}X\]
which can be probabilistically interpreted in the context of $\alpha-$stable
distributions: a distribution $f_{\alpha}$ is $\alpha-$stable if,
for $X_{1}$ and $X_{2}$ independent with distribution $f_{\alpha}$,
the linear combination\[
a_{1}X_{1}+a_{2}X_{2}\sim\left(\vert a_{1}\vert^{\alpha}+\vert a_{2}\vert^{\alpha}\right)^{\frac{1}{\alpha}}X,\]
where $X$ follows again distribution $f_{\alpha}.$ Thus, a Gaussian
distribution is $\alpha-$stable with $\alpha=2.$ The result of Thm.1
can be viewed as follows: $q-$Gaussians are not $\alpha-$stable
(unless $q=1$ which corresponds to the Gaussian case $\alpha=2$);
however, their scaling behavior is close to the Gaussian $\alpha=2$
case, except for the fact that the scaling variable $a_{1}\circ a_{2}$
includes an additional random term $\lambda$.

\section{Geometric interpretation}

Geometrically, the Gaussian scaling factor $\sqrt{a_{1}^{2}+a_{2}^{2}}$
can be interpreted, according to Pythagoras' theorem, as the length
of the hypotenuse of a right triangle with sides of lengths $\vert a_{1}\vert$
and $\vert a_{2}\vert$. The $q-$Gaussian case corresponds to a triangle
for which the angle between $\vert a_{1}\vert$ and $\vert a_{2}\vert$,
let us call it $\phi$, fluctuates around rectangularity.

\begin{figure}[h]
\begin{centering}
\includegraphics[scale=0.4]{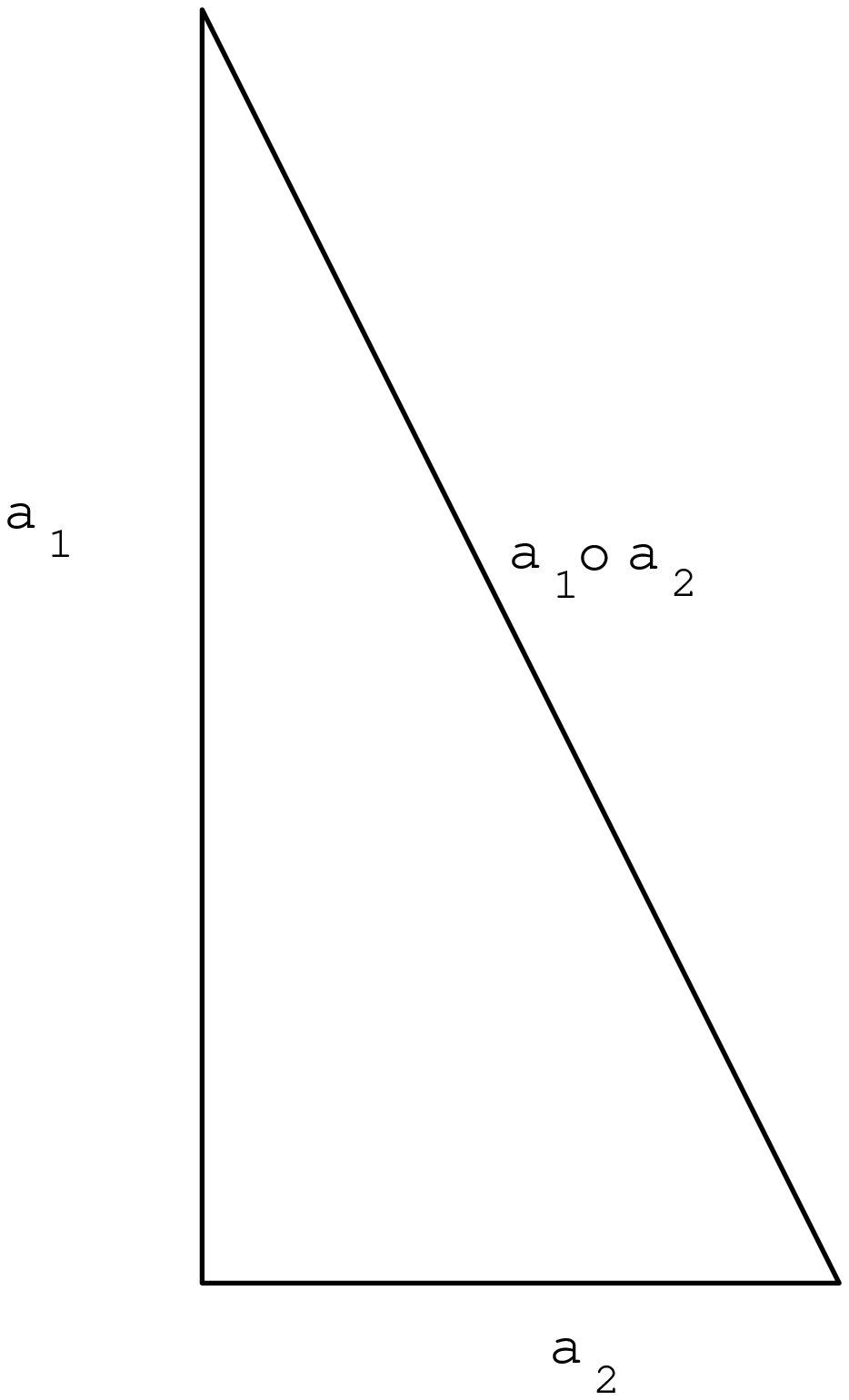}\includegraphics[scale=0.4]{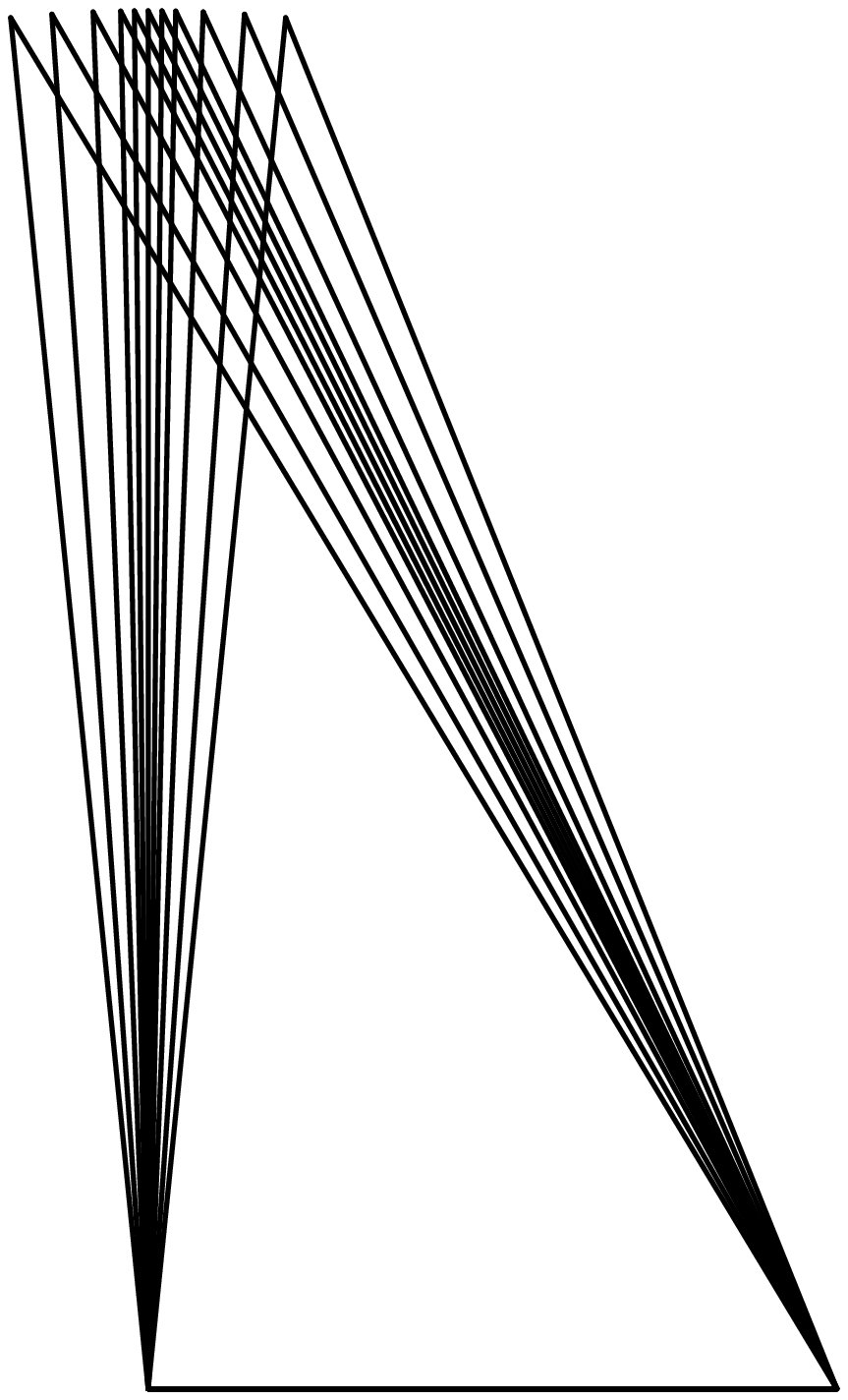}
\end{centering}
\caption{the geometric interpretation of $a_{1}\circ a_{2}$ in the Gaussian
case ($q=1$ left); in the $q-$Gaussian case (left), $a_{1} \circ a_{2}$ is randomly chosen as one of the hypothenuses represented, the angle $\phi$ between sides $a_1$ and $a_2$ being distributed as shown on Figure \ref{fig:fig1}}
\end{figure}

\begin{center}
\begin{figure}
\centering{}\includegraphics[scale=0.35]{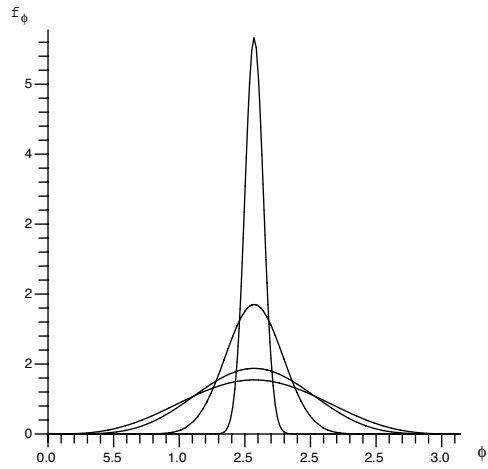}
\caption{\label{fig:fig1}the distribution of angle $\phi$ for
values of the parameter $q_{\lambda}=0.99,\,0.9,\,0.5$ and $0.1$
(top to bottom).}
\end{figure}
\end{center}

The distribution of the angle $\phi$ where $\lambda=-\cos\phi$ is
given by 
\[
f_{\phi}\left(\phi\right)=\frac{\Gamma\left(\nu+1\right)}{\Gamma\left(\nu+\frac{1}{2}\right)\Gamma\left(\frac{1}{2}\right)}\sin^{2\nu}\phi,\,\,0\le\phi\le\pi,\,\,\nu=\frac{1}{1-q_{\lambda}}+\frac{1}{2}.
\]
This distributions is shown in Figure \ref{fig:fig1} for values of the parameter $q=0.99,\,0.9,\,0.5$
and $0.1$ (top to bottom).

We remark that this distribution is symmetric around the angle $\phi=\frac{\pi}{2}$
and that, as $q\to1,$ the angle $\phi$ becomes deterministic and
equal to $\frac{\pi}{2}$. Further, the usual scaling law for Gaussian
distributions (\ref{eq:Gaussiancase}) is recovered.

\subsection{An optical analogy}

We remark that formula (\ref{eq:algebra}) exhibits a close
resemblance with the interference formula for the amplitude of the
superposition of two optical beams. Interferometric optical
testing is based on these phenomena of interference. Two-beam
interference is the superposition of two waves, such as the
disturbance of the surface of a pond by a small rock encountering
a similar pattern from a second rock. When two wave crests reach
the same point simultaneously, the wave height is the sum of the
two individual waves. Conversely, a wave trough and a wave crest
reaching a point simultaneously will cancel each other out. Water,
sound, and light waves all exhibit interference. A light wave can
be described by its frequency, amplitude, and phase, and the
resulting interference pattern between two waves depends on these
properties, among others. Our present interest lies in the
two-beam interference equation. It gives the irradiance $I$
\cite{irrad} for monochromatic waves of irradiance $I_1,$ and
$I_2$ in terms of the phase difference $\Delta_{1,\,2}$ expressed
as $\cos{\phi}= \cos{(\phi_1-\phi_2)}$. We have

\[ I=I_{1}+I_{2}+2\sqrt{I_{1}I_{2}}\cos{\phi},\] and, in
terms of the $A-$amplitudes $I=A^2$,
\[ A^2=A_{1}^2+A_{2}^2+2A_{1}A_{2}\cos{\phi}.\]
If the emission of the two beams could be so arranged that the
phase difference becomes random \cite{pra08,lukin,french}, this
physical analogy would be exact.

\subsection{Study of the composition law $\circ$}

The composition law \[
a_{1}\circ a_{2}=\sqrt{a_{1}^{2}+a_{2}^{2}+2\lambda a_{1}a_{2}}\]
 has been studied in \cite{Kingman}, in the more general case where
$a_{1}$ and $a_{2}$ are independent, positive random variables.
The associativity result is as follows
\begin{thm}
\cite[p.18 thm.1]{Kingman} The composition law $\circ$ is associative
if and only if either
\[
a_{1}\circ a_{2}=\sqrt{a_{1}^{2}+a_{2}^{2}}
\]
or
\[
a_{1}\circ a_{2}=\vert a_{1}\vert+\vert a_{2}\vert
\]
or
\begin{equation}
a_{1}\circ a_{2}=\sqrt{a_{1}^{2}+a_{2}^{2}+2\lambda a_{1}a_{2}}\label{eq:a1a2}
\end{equation}
where $\lambda\sim G_{q}\left(0,1\right)$ for some $q\ge0.$
\end{thm}
This theorem can be interpreted as follows: the only cases where the
composition law $\circ$ is associative is
\begin{enumerate}
\item the Gaussian case ($q=1, \, q_{\lambda}=1$)
\item the Cauchy case ($q=2$)
\item the present $q-$Gaussian case with $0 \le q < 1$
\end{enumerate}
It is thus a remarkable property that the only cases of associativity of this composition law correspond to the whole range of $q-$Gaussian distributions with $0 \le q \le 1$ or to the Cauchy case $q=2$.
We note moreover that in the limit case $q=0,$ $a_{1}\circ a_{2}$
in (\ref{eq:a1a2}) reduces to a Bernoulli random variable that takes
values $a_{1}+a_{2}$ and $a_{1}-a_{2}$ with probability $\frac{1}{2}.$

\subsection{A Central Limit Theorem for the composition law $\circ$}

In the same spirit as the central limit theorem for the usual addition,
a central limit theorem exists for the composition law $\circ$. Before
we give its rigourous expression as established in \cite{Kingman},
let us look at a special case of it based on the theory of superstatistics.
Let us consider the random walk\[
S_{n}=\frac{1}{\sigma\sqrt{n}}\sum_{i=1}^{n}X_{i}\]
where $X_{i}$ are independent $q-$Gaussian random variables with
same variance $\sigma^{2}$ and where $q<1;$ since all $X_{i}$ have
finite variance, the usual Central Limit theorem applies and \[
S_{n}\to\mathcal{N}\left(0,1\right).\]
But by theorems (\ref{thm:associative}) and (\ref{thm:generalsum}),
we have also\[
S_{n}=\left(\bigotimes_{i=1}^{n}\frac{1}{\sigma\sqrt{n}}\right)X\]
with notation\[
\bigotimes_{i=1}^{n}a_{i}=a_{1}\circ\dots\circ a_{n}\]
 where $X$ is $q-$Gaussian with the same nonextensivity parameter
$q.$ Since, by the superstatistics theory, a Gaussian random variable
$G$ can be decomposed as 
\[
G=\chi_{p}X
\]
where $\chi_{p}$ is chi-distributed with $p$ degrees of freedom,
we deduce that the following limit
\[
\bigotimes_{i=1}^{n}\frac{1}{\sigma\sqrt{n}} =
\frac{1}{\sigma\sqrt{n}} \bigotimes_{i=1}^{n} \,1\underset{n\to+\infty}{\longrightarrow}\chi_{p}
\]
should hold. But this result is easy to check at least under hypothesis
(H): in this case, \[
S_{n}=r_{n}X\]
where \[
r_{n}=\vert\sum_{i=1}^{n}U_{i}\vert\]
where $U_{i}$ are independent and uniformly distributed on the sphere
$\mathcal{S}_{p-1}.$ By the Central Limit Theorem, $r_{n}\to\vert G\vert$
where $G$ is a Gaussian vector in $\mathbb{R}^{p}$; hence $r_{n}$
converges indeed to a $\chi$ distributed random variable with $p$
degrees of freedom.

It turns out that a much more general result holds, namely
\begin{thm}
\cite{Kingman} If $\left\{ a_{i}\right\} $ are positive, independent
and identically distributed random variables with variance $\sigma$,
then the composition
\[
\frac{1}{\sigma\sqrt{n}}\bigotimes_{i=1}^{n}a_{i}\underset{n\to+\infty}{\longrightarrow}\chi_{p}
\]
 where $\chi_{p}$ is a chi-distributed random variable\footnote{we recall that parameter $p$ enters the picture through the distribution of the random variable $\lambda$ included in the composition law}with parameter  
$p.$
\end{thm}
This result can be considered as a central limit theorem for the algebra $\circ$ defined by (\ref{eq:algebra}).

\section{Conclusions}

In this work we have uncovered interesting geometric aspects that
underlie the sum $Z$ of two stochastic variables $a_{1}X$ and
$a_{2}X_{2}$ ($a_{1}$, $a_{2}$ are scalars and $X_{1}$, $X_{2}$
are $n-$variate vectors). The alluded geometry becomes operative
when the two variables  are governed by $q-$Gaussian probability
distributions with $q<1$. We found that its sum $Z$ turns out to be
$q-$Gaussian with same nonextensivity parameter $q$ multiplied by
an independent random factor $a_{1}\circ a_{2}$. In turn,
the random factor can be described as a random and symmetric
mixture of the two constants $a_{1}$ and $a_{2}$, the random
factor involved following itself a $q-$Gaussian distribution.


\begin{thebibliography}{5}
\bibitem{uno}C. Tsallis, \textit{Introduction to Nonextensive Statistical
Mechanics: Approaching a Complex World} (Springer-Verlag, New York,
2009)

\bibitem{umarov}S. Umarov, C. Tsallis, S. Steinberg, On a q-Central
Limit Theorem Consistent with Nonextensive Statistical Mechanics,
Milan Journal of Mathematics, 76-1, 307-328, 2008

\bibitem{Fang}K.-T. Fang , S. Kotz, K. W. Ng, Symmetric Multivariate
and Related Distributions, Chapman \& Hall/CRC, 1989

\bibitem{Watson}G.N. Watson, A treatise on the theory of Bessel functions,
second edition, Cambridge, (1944)

\bibitem{Kingman}J.F.C. Kingman, Random walks with spherical symmetry,
Acta Math. 109, 11-53, (1965).

\bibitem{irrad} Irradiance, radiant emittance, and radiant exitance
are radiometry terms for the power of electromagnetic radiation at
a surface, per unit area. The term irradiance is used when the
electromagnetic radiation is incident on the surface.

\bibitem{pra08} T. Setala, A. Shevchenko, M. Kaivola, and A. T. Friberg,
 Phys. Rev. A {\bf 78}, 033817 (2008).

\bibitem{lukin} V. P. Lukin, Kvantovaia Elektronika (Moscow), {\bf 7}, 1270 (1980).

\bibitem{french} Z. H. Gu, H. M. Escamilla, E. R. Mendez, A. A.  Maradudin, J. Q. Lu,
T. Michel, M. Nieto-Vesperinas, Applied Optics {\bf 31}, 5878
(1992).

\end{thebibliography}
\end{document}